%% file: IC-SS-AA.TEX
\documentclass[preview]{elsarticle}
\input{SETI-headers}  

\usepackage{mathpazo}
\usepackage[scaled=0.87]{berasans}
\usepackage[scaled=0.87]{beramono}
\normalfont
\usepackage[T1]{fontenc}
\usepackage{textcomp}

\newcommand{\pfa}{P_{\text{FA}}}
\newcommand{\pd}{P_{\text{D}}}

\title{Interstellar Communication: The Case for Spread Spectrum} 

\author[usa]{David G Messerschmitt}

\address[usa]{Department of Electrical Engineering and Computer Sciences, 253 Cory Hall, University of California, Berkeley, California 94720-1770, USA, messer@eecs.berkeley.edu, 1-925-465-5740}

\begin{document}

\begin{abstract}

Spread spectrum, widely employed in modern digital wireless terrestrial radio
systems, chooses a signal with a noise-like character
and much higher bandwidth than necessary.
This paper advocates spread spectrum modulation for interstellar
communication, motivated by 
robust immunity to radio-frequency interference (RFI)
of technological origin in the vicinity of the receiver while
preserving full detection sensitivity in the presence of
natural sources of noise.
Receiver design for noise immunity alone
provides no basis for choosing a signal with any specific character,
therefore failing to reduce ambiguity.
By adding RFI to noise immunity as a design objective,
the conjunction of choice of signal (by the transmitter) together with
optimum detection for noise immunity (in the receiver) leads through simple
probabilistic argument to the conclusion that the signal should possess
the statistical properties of a burst of white noise,
and also have a large time-bandwidth product.
Thus spread spectrum
also provides an implicit coordination between transmitter
and receiver by reducing the ambiguity as to the signal character.
This strategy requires the receiver to guess
the specific noise-like signal, and it is contended that this is feasible
if an appropriate pseudorandom signal is generated algorithmically.
For example,
conceptually simple algorithms like the binary expansion of
common irrational numbers like $\pi$ are
shown to be suitable.
Due to its deliberately wider bandwidth,
spread spectrum is more susceptible to dispersion
and distortion in propagation through the interstellar medium,
desirably reducing ambiguity
in parameters like bandwidth and carrier frequency.
This suggests a promising new direction in
interstellar communication using spread spectrum modulation techniques.

\end{abstract}


\maketitle

\section{INTRODUCTION}

This paper addresses the design of an end-to-end digital interstellar
communication system that can be used to exchange information among
intelligent civilizations.
As advocated in \citeref{583}, we employ communication's engineering
principles to achieve an implicit form of coordination.
This strategy leverages the physical impairments that are unavoidable in an end-to-end
interstellar digital communication system, and presumably known to both transmitter and receiver,
and bases design decisions on mitigation of those impairments.
Where those design decisions are based on mathematical optimization, the
transmitter and receiver designers will necessarily arrive at compatible designs.
Where optimization is not possible, the receiver must apply judgement and/or
systematically search over the possibilities,
minimizing ambiguity by keeping the design as simple as possible
within the constraints of acceptable cost and performance.

There are two distinct phases to establishing information transfer \citeref{583}:
(a) receiver discovery of the existence of a signal followed by 
(b) extraction of information from that signal.
This is rendered more challenging by the impossibility of advance coordination
between transmitter and receiver, and discovery is the more challenging problem
and also represents the dominant cost and effort for a receiving civilization.
Although this paper addresses both discovery and communication from
an end-to-end perspective,
past efforts have addressed transmission and reception as primarily separate issues,
and in the case of the receiver (with the exception of \citereftwo{581}{577}) have focused
on discovery of signals designed to attract attention but not bear information.
Some entryways into the extensive literature on this topic 
can be found in \citerefsix{120}{129}{131}{587}{585}{584}.
Longstanding programs try to discover signals of
technological origin \citeref{592} (called ``SETI'' \citeref{120}).
Other efforts have transmitted signals with an embedded message
(called ``METI'' \citeref{586}), including
 the ``Arecibo message'' \citeref{591} in 1974 and more recently
``Cosmic Call'', ``Teen Age Message'' and ``A Message from Earth'' \citeref{585}.
The transmitter side has also been studied from a cost perspective \citereftwo{568}{569}.

The end-to-end design of interstellar communication systems is relevant to both SETI and METI.
Our focus is on radio frequencies, digital modulation
(although analog modulation has also been experimented with \citeref{585} in METI),
and communication of an information-bearing message \citeref{584}.
Many SETI projects have followed the Cyclops Report \citeref{142} more than four decades ago,
which recommended searching for extremely narrow bandwidth signals.
Several factors have changed since then.
First, the explosion in
terrestrial wireless multi-user communications systems
has improved our understanding of radio communications
in an RFI-dominated environment \citerefthree{191}{159}{193}.
This has strongly shifted the design of radio systems toward
spread spectrum techniques, in which the signal bandwidth is much
greater than dictated by the supported information rates \citereftwo{157}{160}.
Second, there has been great progress in understanding the 
radio propagation properties of the interstellar medium (ISM) based
in large part on pulsar astronomy \citeref{197}.
Third, dramatic advances in electronics technology
open up new possibilities in our own search capabilities.

Here we consider the specific design criterion of maximizing the 
probability of discovering an information-bearing signal and the correct detection of its
embedded information content.
This criterion is appropriate for a transmitter constrained in transmit power
and seeking to maximize the distance over which a message signal can be discovered
and its information content extracted.
For a given transmit power this criterion also maximizes the 
number of stellar systems wherein
discovery is feasible, 
and in this sense the probability that the signal will be discovered
by someone somewhere.
Even if the details of the transmit signal are completely known to the receiver,
there are four primary impairments beyond the control of transmitter and receiver designers
that impinge on discovery and detection:
natural noise sources, radio-frequency interference (RFI) in the receiver vicinity,
radio propagation effects \citeref{582} through the interstellar medium (ISM),
and Doppler due to relative accelerations.
This paper focuses on the first two, and subsequent papers
(starting with \citeref{572}) will address the remaining.

RFI is a challenge for SETI because its technological origin
mimics the interstellar signal being sought.
An ongoing trend on earth and its immediate environs
is ever-increasing local RFI.
It is possible that a transmitter designer (which is likely to be more
technologically mature than us) has experienced a similar
growth in RFI and also has developed an understanding of RFI mitigation
in the context of its own multi-user wireless communication systems.
If so, the transmitter designer may view RFI mitigation
in the receiver environment as a significant design objective.
We demonstrate in this paper that this circumstance would be very fortunate,
for two reasons.
First, signal choice by the transmitter taking RFI into account can
strongly mitigate the adverse effects of RFI on
our own SETI  observations, directly addressing our growing challenge.
Second, considerations of signal acquisition in the presence of RFI allow us to
infer very specific characteristics of an information-bearing signal, considerably beyond
what design for noise (as pursued in \citeref{142} and elsewhere) can infer.
This establishes principles allowing us to infer
what type of signal may have been chosen by the transmitter.
In contrast,
even recent attempts at applying communication theory to SETI have emphasized
noise sensitivity \citerefsix{196}{578}{140}{581}{127}{602},
and are as a result unable to infer any concrete characteristics of the signal
even as many advocate the use of greater bandwidth.
The propagation characteristics of the ISM introduce further
 constraints on the bandwidth and carrier frequency of the transmit signal, which
 desirably limits the ambiguity in these parameters for both
 transmitter and receiver \citeref{572}.
Others have emphasized such astrophysical effects in trying to infer 
signal characteristics \citeref{123}.

The focus of this paper is on the choice of a transmitted signal,
which directly parallels the receiver's challenge 
of anticipating what type of signal to expect.
In this we take the perspective of a transmitter designer, because
in the absence of explicit coordination it is the
transmitter, and the transmitter alone, that chooses the signal.
This is significant because the transmitter designer possesses far less information about the receiver's
environment than the receiver designer, due to both distance (tens to hundreds of light-years)
and speed-of-light delay (tens to hundreds of years).
While the receiver design can and should take into account all relevant characteristics
of its local environs and available resources and technology,
in terms of the narrower issue of what type of signal to
expect the receiver designer must rely exclusively on the perspective of the transmitter designer.

The primary result of this paper is as follows.
The transmitter designer should assume that the receiver employes
the optimum detection algorithm for white Gaussian noise, but assume
nothing specific about the RFI environment of the receiver since nothing is presumably known.
The transmitter can, however, explicitly design a transmit signal that minimizes the
effect of RFI on the receiver's discovery and detection probabilities
in a robust way; that is, in a way that provides a constant immunity
regardless of the nature of the RFI.
It is shown that the resulting immunity increases with the
product of time duration and bandwidth,
and that the signal
should resemble statistically a burst of white noise.
Intuitively this is advantageous because RFI resembles
such a signal with a likelihood that decreases exponentially with time-bandwidth product.
Both a transmitter and receiver designer using this optimization criterion
and employing the tools of elementary probability theory will arrive
at this same conclusion.
Although the context is different,
variations on this principle inform the design of many modern
widely deployed terrestrial digital wireless communication systems,
so this has been extensively tested in practice and is likely to have a prominent place
in the technology portfolio of an extraterrestrial civilization as well.

\section{DESIGN APPROACH}

This section qualitatively describes an end-to-end digital communication
system including counteracting noise and RFI.
Section \ref{sec:consequences} brings to bear a relevant mathematical principle drawn from probability theory.

\subsection{Modulation}

An intermediate step of modulation (\ref{sec:infobear})
accommodates the analog nature of a radio channel.
A common case is pulse-amplitude modulation (PAM), in which a sequence 
of information-bearing symbols $\{ A_k \}$ , each drawn from a finite constellation
(finite set of discrete points), is used to adjust the amplitude of a
sequence of pulses with shape $h(t)$,
\begin{equation}
\label{eq:pam}
y(t) = \sqrt{\energy_{h}} \cdot \sum_k A_k \, h(t - kT) \,.
\end{equation}
It is assumed that
 $h(t)$ is a unit-energy pulse waveform, 
 $1/T$ is the symbol rate, and
 $\energy_{h}$ is the energy per symbol.
 If the receiver can distinguish which $A_{k}$ was transmitted,
 it can extract the information being conveyed.
 The average power is $\power_{s} = \energy_{s}/T$, and if
 it is desired to reduce $\power_{s}$ (at the expense of lower
 information rate) this can be done by increasing $T$.
The radio propagation must be passband with some carrier frequency
$f_{c}$, and as described in \ref{sec:infobear} this allows
the baseband-equivalent signal $y(t)$ in \eqref{eq:pam} to be complex-valued.\footnote{
At passband $A_{k}$ is interpretated
 as the carrier amplitude and phase for symbol period $k$.}
This paper focuses on the choice of $h(t)$ and its implications
to the extraction of $\{ A_k \}$ from $y(t)$ in the face of noise and RFI.
 
\subsection{Detection in noise}

Thermal noise is ubiquitous in nature; at radio frequencies, it can be
modeled at baseband as an additive white Gaussian noise $N(t)$. 
Here we focus on the detection of  representative one symbol $A_0$.
If $h(t)$ is known to (or can be guessed by) the receiver,
the optimum receiver processing \citeref{43} is
a matched filter (MF),
\begin{equation}
\label{eq:ctmf}
Z = \int \left( y(t) + N(t) \right) \, \cc h (t) \, dt = \sqrt{\energy_{s}} \, A_{0} + \int N(t) \, \cc h (t) \, dt \,.
\end{equation}
For \eqref{eq:ctmf} to be valid the time translates of $h(t)$ must also be orthogonal,
\begin{equation}
\label{eq:ott}
\int h(t-kT) \, \cc h (t) \, dt = \delta_{k}
\ \ \ \ \text{or} \ \ \ \
\sum_m \, \left| \, H \left( f + \frac{m}{T} \right) \, \right|^{\,2} = \frac{1}{T} \,.
\end{equation}
The decision variable $Z$ in \eqref{eq:ctmf} can be used for two purposes.
During communication, $Z$ can be applied to a complex-valued slicer
that takes account of the symbol constellation to make a decision on which
$A_{0}$ was transmitted.
During discovery the symbol constellation is not known,
but as long as $h(t)$ is known or can be guessed  $| Z |$ (an estimate of
$\sqrt{\energy_{s}} \, | A_{0} |$) can be applied to a real-valued threshold to decide
whether or not a signal is present.
Thus, the MF can be used for discovery based on single-symbol energy, the approach used here.
If more assumptions are made about the symbol constellation and any
redundancy in successive symbols, then the approaches here can be modified to
base discovery on multiple symbols,
thereby capturing more signal energy (see \citeref{577} for an example).

The statistics of the noise term in $Z$ in \eqref{eq:ctmf}  do not depend on $h(t)$, including its
time duration or bandwidth.\footnote{
The statement in \citeref{131} that ''spreading the bandwidth makes detection more difficult'' 
does not apply to a MF.}
As a result, design issues related to noise offer no clue as to what $h(t)$
should be chosen by the transmitter.
This is paradoxical since the MF also assumes that
the receiver knows or can guess $h(t)$.

\subsection{Radio-frequency interference}
\label{sec:rfi}

Radio astronomy is increasingly impacted by RFI with increasing
receiver sensitivity and growing sources of RFI \citereftwo{604}{608}, and many
techniques have been successfully applied to RFI mitigation \citeref{606}.
This growing problem has even stimulated extreme measures like planning for
observations on the far side of the moon \citereftwo{198}{199}.
Many of the same instruments, challenges, and techniques can be shared with
interstellar communication, including those listed in Table \ref{tbl:rfi}.
Interstellar communication has the special challenge that RFI and the
interstellar communication signal share a technological origin, and
may even be communication signals using similar modulation techniques.
In the discovery phase,
RFI creates two very different anomalies: False alarms due to RFI being
mistaken for signal, as well as misses due to RFI overpowering a signal of interest.
A special opportunity applicable only to communication
(technique V. in Table \ref{tbl:rfi})
is to engineer the signal structure to make it distinguishable
from RFI.
This approach is familiar in terrestrial and space-based communication systems,
since they also have to deal with RFI and the challenge of distinguishing one radio-based
communication from another.
Gains arising from this approach are additive to gains from other mitigation measures,
and unlike many techniques used in astronomy they
can be effective against RFI with time, spectral, and spatial overlap with the communication signal,
reducing both false alarms and misses.
A disadvantage is the required cooperation of the transmitter.

\doctable
	{\small}
	{rfi}
	{
	Some techniques for dealing with RFI borrowed from radio astronomy
	and their efficacy for interstellar communication.
	It is assumed  the receiving antenna is highly directional
	and that RFI originates from a point source.
	The terminology is DOA = ''direction of arrival`` and SOI = ''signal of interest''.
	}
	{l p{1.5cm} p{6cm} p{6cm}}
	{
	&&\textbf{Technique}&\textbf{Interstellar communication}\\
	\toprule
	\addlinespace[3pt]
	I.&Spatial \newline rejection&
	Situate receiving antennas in quiet locations.
	Reduce spurious antenna sidelobes by antenna design.
	Space-based RFI sources not in the SOI's DOA are rejected.&
	Effective, although an RFI source in the DOA is difficult to distinguish
	from a SOI since they are both of technological origin.
	\\
	\addlinespace[3pt]
	\cmidrule{3-4}
	II.&Temporal \newline rejection& 
	If an RFI source is transient, observe during periods it is absent.
	Most space-based sources of RFI (even geostationary satellites) are
	in motion relative to a SOI's DOA.
	RFI may exhibit distinguishing periodicities \citeref{609}.&
	The SOI will likely be transient (due to stellar scanning, random scintillation, etc.),
	but its periodicity can be chosen to be non-typical of near-space RFI.
	\\
	\addlinespace[3pt]
	\cmidrule{3-4}
	III.&Frequency \newline rejection& 
	If an RFI source is frequency-specific, observe at other frequencies.
	RFI may have a recognizable narrowband component signature.
	Some frequency bands are reserved for passive use (no transmissions).&
	A discovery search cycles through frequency, and
	only RFI overlapping the SOI bandwidth can
	cause false alarms or misses.
	Passive frequency bands offer less
	 protection since a remote
	 transmitter is not cognizant of their location.
	\\
	\addlinespace[3pt]
	\cmidrule{3-4}
	IV.&Spatial \newline filtering& 
	For multiple-element antennas, 
	take advantage of high or low correlation
	of RFI across elements and different DOA's of RFI and SOI \citeref{607}.
	Employ array signal processing, such as RFI tracking, adaptive interference cancellation, etc.&
	Effective, with the same caution as in I.
	\\
	\addlinespace[3pt]
	\cmidrule{3-4}
	V.&Signal \newline character&
	SOI's may have a special structure (like the periodicity of pulsars \citeref{212}) that distinguishes them from
	RFI or makes intermediate ISM impairments evident \citeref{605}.&
	With the cooperation of the transmitter,
	engineer the SOI's structure to make it more readily distinguishable from RFI.
	\\
	\addlinespace[3pt]
	\bottomrule
	}

This paper is dedicated to investigating this signal structure approach.
To address the effect of RFI on communication, observe that
the matched filter of \eqref{eq:ctmf} does not take account of RFI.
If further knowledge of the
characteristic of RFI were incorporated into the detection, then sensitivity could be improved.
However, since the transmitter designer presumably has no specific knowledge of  RFI
local to the receiver,
that designer should choose $h(t)$ assuming only that the noise-optimal MF is used for detection.
Even subject to this limitation,
the transmitter designer can assist the receiver through the choice
 of $h(t)$, and that observation provides specific and 
credible information about the nature of $h(t)$
that a receiver should seek.

Suppose the baseband-equivalent RFI at the receiver is $r(t)$.
The transmitter designer should take the stance that $r(\cdot)$ is an unknown waveform,
and even be unwilling to assign a statistical distribution (which assumes that some
$r(\cdot)$'s occur more frequently than others).
The transmitter can anticipate the magnitude of the
contribution of RFI to $Z$ in \eqref{eq:ctmf} as
\begin{equation}
\label{eq:cti}
\left| \, Z \, \right| ^{2} = 
\left| \, \int_{0}^{T} r (t) \, \cc h (t) \, dt \, \right|^{\,2} =
\left| \, \int_{0}^{B} R (f) \, \cc H (f) \, df \, \right|^{\,2}
\le \energy_{r} (T,B) \,,
\end{equation}
where it is assumed that $h(t)$ is simultaneously bandlimited to $f \in [0,B]$ Hz
(so the passband representation is bandlimited to $|f| \in [f_{c}, f_{c}+B]$)
and time-limited to $t \in [0,T]$ seconds.\footnote{
This arbitrary choice of $f_{c}$ (so that the baseband signal has a one-sided spectrum) emphasizes
that $h(t)$ is complex-valued.
The ideal of simultaneous time and bandwidth limitation can be
achieved with increasing accuracy \citeref{599} as $K = BT \to \infty$.}
This RFI term corrupting the decision can be small or large in magnitude
depending on how $r( \cdot )$ correlates (or doesn't correlate) with signal $h(t)$,
but its largest value is $\energy_{r} (T,B)$, the energy of $r(t)$
confined to  $t \in [0,T]$ and $f \in [0,B]$.

With respect to RFI, the receiver gain $G \ge 1$ is defined as the
ratio of the input RFI energy to the RFI energy appearing at the decision threshold.
The larger $G$ the better the receiver has suppressed RFI
before decision making.
From \eqref{eq:cti} there are two obvious components in $G = FG \times PG$.
The \emph{filtering gain} $FG = \energy_{r} (\infty,\infty) / \energy_{r} (T,B)$ 
is the reduction in energy due to confinement to
$t \in [0,T]$ and $f \in [0,B]$
(an example of techniques II. and III. in Table \ref{tbl:rfi}).
The \emph{processing gain} $PG$ (an example of IV. and V.)
is the reduction in energy below $ \energy_{r} (T,B)$
due to a less-than-maximum correlation 
between $r(t)$ (after confinement in bandwidth and time) and $h(t)$.
Clearly the worst case is $r(t) = \sqrt{\energy_{r}} \, h(t)$, for which
$G = FG = PG = 1$.
Not surprisingly,
a receiver cannot distinguish $h(t)$ from an identical RFI
(within a constant amplitude and phase shift).

\section{ISOTROPIC PRINCIPLE AND ITS CONSEQUENCES}
\label{sec:consequences}

An appropriate choice of $h(t)$
can insure that $PG \gg 1$ consistently in a statistical sense.
One key is to choose $K = B T \gg 1$.
This is called \emph{spread spectrum} because
$B$ is larger than the minimum
required, which is $B \ge 1/T$ from \eqref{eq:ott}
(this is called the Nyquist criterion \citeref{43}).
Thus large $PG$ results in part from choosing a bandwidth $B$ much larger than the
minimum required for a given symbol rate $T$.
This will follow from the
 \emph{isotropic principle} (IP), which explains both
the optimality of the MF in \eqref{eq:ctmf} and the relationship between bandwidth and $PG$.

\subsection{Signal degrees of freedom}

Assume that $h(t)$ equals a weighted linear combination of $K = BT$
orthonormal functions, implying that
$h(t)$ actually falls in a $K$ dimensional space.
Common examples include
\begin{align}
\label{eq:oseries1}
h(t) &= \frac{1}{\sqrt{B}} \, \sum_{k=0}^{K-1} h \left( \frac{k+ 1/2}{B} \right) \, \phi \left( \frac{t - k - 1/2}{B} \right) \\
\label{eq:oseries2}
h(t) &= \frac{1}{\sqrt{T}}  \, \sum_{k=0}^{K-1}  H_{k} \, e^{\,i \,2 \pi \,k t /T}
\end{align}
The first form in \eqref{eq:oseries1} is the sampling theorem, which uses 
time translates of a appropriate unit-area interpolation function $\phi (\cdot)$
to generate an $h(t)$ that
 is band limited to $f \in [0,B]$ and approximately time limited to $t \in [0,T]$.
The second form in \eqref{eq:oseries2} is the Fourier series, 
which generates an $h(t)$ that is time-limited to $t \in [0,T]$
(with periodic extension) and approximately band limited to $f \in [0,B]$.
Both expansions illustrate that $B$ and $T$ can be chosen
independently (as long as $B T \ge 1$), and this will
not affect the detection noise immunity
as long as $\energy_{s}$ is fixed.\footnote{
The vague notion of waveforms that are both band limited and time limited can be
made precise by choosing a set of orthonormal functions as these examples illustrate
and then defining a linear span of these functions \citeref{599}.
The assertion in \citeref{142} that increasing $T$ in the interest of reducing signal power
''narrows the signal spectrum'' and as a consequence 
''we would expect interstellar contact signals to be highly monochromatic''
is unjustified because $B$ can be chosen independently of $T$.}
Increasing $T$ does reduce both $\power_{s}$ and information rate.

An orthonormal representation converts the problem of designing $h(t)$ 
to the choice of a $\esv h \in \cn{K}$
(Euclidean space of $K$-dimensional complex-valued column vectors).
With respect to any basis, the reception after band limiting to
$f \in [f_c, \, f_c + B]$ and translation to baseband $f \in [0, \, B]$
can be represented by a random $\esv Y \in \cn{K}$ as
\begin{equation}
\label{eq:reception}
\esv Y =
\sqrt{\energy_h} \,  e^{\,i \, \theta} \cdot \esv h + \esv N_1 + \sqrt{\energy_r} \cdot \esv r \,,
\end{equation}
where
$\energy_h$ is an unknown signal energy,
$\theta$ is an unknown carrier phase,
$\esv h$ is a known unit-energy pulse,
$\esv N_1$ is a random noise vector,
$\energy_r$ is an unknown RFI energy, and
$\esv r$ is an unknown unit-energy RFI vector.
The energy of any $\esv x \in \cn{K}$ is $\| \esv x \|^{\,2} = \hc{\esv x} \esv x$, where
 $\hc{\esv x}$ denotes the transpose and conjugate of $\esv x$.

Any $\esv h$ points in a single direction in $\cn{K}$,
and $K$ specifies the number of complex numbers required to specify that direction,
or equivalently the dimensionality of the space containing $\esv h$.
Thus $K$, which we call the signal \emph{degrees of freedom} (DOF), 
quantifies the ambiguity in the signal direction.
Even $\esv h = \trans{[1,\;0, \dots ,\;0]}$ has a DOF of $K$; 
it just happens that $K-1$ of those
degrees have been chosen to be zero.

\subsection{Completely random vectors and the isotropic principle}

\emph{Completely random} (CR) vectors play a fundamental role in both
noise analysis and signal design.
A random $\esv V \in \cn{K}$ 
is fully specified by $2K$ real-valued random variables
$\left\{ \real{V_k} , \, \imag{V_k} , \, 1 \le k \le K  \right\}$
with some joint probability distribution.
A Gaussian CR $\esv V$ is
zero-mean ($E [ \esv V ] = \esv 0$), and all of its $2K$ constituent random variables
are jointly Gaussian, identically distributed with variance $\sigma^{\,2}/2$, and 
statistically independent.
These conditions can be relaxed as shown in Table \ref{tbl:flavorCR}
by eliminating the Gaussian and/or independence conditions.

\doctable
	{\small}
	{flavorCR}
	{
	Three flavors of complete randomness (CR) for a random $\esv V \in \cn{K}$.
	}
	{p{3cm}p{11cm}}
	{
	\toprule
	\addlinespace[3pt]
	White CR& 
	All $2K$ real and imaginary parts are zero-mean, mutually uncorrelated,
	and have the same variance $\sigma^{\,2}/2$.
	\\
	\addlinespace[3pt]
	\cmidrule{2-2}
	Independent CR&
	All $2K$ real and imaginary parts are zero-mean and statistically independent. 
	Since independent random variables are uncorrelated, independent CR implies white CR.\\
	\addlinespace[3pt]
	\cmidrule{2-2}
	Gaussian CR&
	White CR, and in addition
	all $2K$ real and imaginary parts have a joint Gaussian distribution.
	Since uncorrelated Gaussian variables are independent,
	Gaussian CR implies independent CR. \\
	\addlinespace[3pt]
	\bottomrule
	}

CR vectors obey an \emph{isotropic principle (IP)} \citeref{574}
that comes in two flavors listed in Table \ref{tbl:flavorIsotropic}.
Roughly speaking, the IP says that statistics of a white CR vector are
the same regardless of what direction in $\cn{K}$ we look
and regardless of what orthogonal basis we choose.
Thus, the energy (variance) of a white CR vector is spread uniformly
across coordinates, and this remains true after any 
 unitary coordinate transformation (such as the Fourier transform).

\doctable
	{\small}
	{flavorIsotropic}
	{
	Two flavors of the isotropic principle (IP) for a random $\esv V \in \cn{K}$,
	where  $\esv u$ is an arbitrary unit vector ($\| \esv u \| = 1$)
	and $\esv U$ is an arbitrary unitary matrix ($\esv U \hc{\esv U} = \esv I$).
	See  \ref{sec:isotropicPrinciple} for derivations and elaboration.
	}
	{p{3cm}p{11cm}}
	{
	\toprule
	\  \newline  Energy \newline isotropic \newline principle
	\newline ($\esv V$ is white CR) & 
	\begin{itemize}
	\item
	The component in any direction $\esv u$ has the same energy
	($E | \hc{\esv u} \esv V |^{\,2} = \sigma^{\,2}$).
	\item
	$\esv V$ after coordinate transformation ($\esv U \esv V$) is also white CR.
	\end{itemize}
	\\
	\cmidrule{2-2}
	\  \newline \  \newline Distribution \newline isotropic \newline principle
	\newline ($\esv V$ is Gaussian CR)&
	\begin{itemize}
	\item
	$\hc{\esv u} \esv V$ is Gaussian, and thus
	$2 \cdot | \hc{\esv u} \esv V |^{\,2} / \sigma^{\,2}$ has a $\chi^{\,2} (2)$ distribution.
	\item
	$\esv U \esv V$ is Gaussian CR.
	\item
	If $\esv V$ is merely independent CR, then under mild conditions the distribution of 
	$\hc{\esv u} \esv V$  approaches Gaussian as $K \to \infty$.
	This implies that the distribution of $2 \cdot | \hc{\esv u} \esv V |^{\,2} / \sigma^{\,2}$ 
	approaches a $\chi^{\,2} (2)$ distribution
	in the limit as $K \to \infty$.
	\end{itemize}
	\\
	\bottomrule
	}

\subsection{Noise}
\label{sec:noise}

At radio frequencies it is usually assumed that the
passband noise from natural sources
(e.g. the background
radiation of the universe and thermal noise in the receiver)
is white and Gaussian, in which case
$\esv N_1$ is Gaussian CR
with variance $\sigma^{\,2}$ per component (\ref{sec:noiseStats}).
Our goal in detector design is to process $\esv Y$ to
yield a real-valued decision variable $Z$ which is applied to a threshold to
distinguish $\energy_h > 0$ (signal plus noise and RFI) from
$\energy_h = 0$ (noise and RFI alone).
Signal is declared present when $Z > \lambda$, 
define a normalized threshold $\tau = \lambda / \sigma$,
and let $\pfa$ be the false alarm probability and $\pd$ be the detection probability.

For the moment assume the absence of RFI ($\esv r = \esv 0$) and
contrast two cases,
\newcommand{\zm}{Z_{\text{M}}}
\newcommand{\ze}{Z_{\text{E}}}
\begin{equation}
\label{eq:dt}
\zm = \left| \, \hc{\esv h} \, \esv Y \, \right|
= \left| \, \sqrt{\energy_h} \, e^{\,i\,\theta} + \hc{\esv h} \, \esv N_1 \, \right|
\ \ \ \ \text{or} \ \ \ \ 
\ze = \left\| \, \esv Y \, \right\|
= \left\| \, \sqrt{\energy_h} \, e^{\,i\,\theta} \, \esv h + \esv N_1 \, \right\|
\end{equation}
where the MF $\zm$ calculates the 
magnitude of the component of reception $\esv Y$
in the direction of unit vector $\esv h$, and the \emph{energy detector} (ED)
$\ze$ calculates the square root of the total energy in $\esv Y$.
The distribution $\zm$ and $\ze$ are determined in 
 \ref{sec:isotropicPrinciple}.
The detector \emph{sensitivity} is defined as the
 $\energy_{h}$ required to maintain fixed $\pfa$ and $\pd$
as $K$ and $\esv h$ change.

For a given $K$,
the sensitivity of neither the MF or the ED depends on the choice of $\esv h$.
It is in the dependence $\energy_{h}$ on $K$ that the MF and ED differ markedly.
For $K = 1$ they are the same, and for $K >1$ the sensitivity of the MF does not change.
This follows immediately from the IP, since the distribution of $\zm$ does not depend
on either $\esv h$ or $K$.
On the other hand, for the ED $\energy_{h}$ must increase by about
10 dB for each $10 \times$ increase in  $K$.
This behavior is explained by Figure \ref{fig:pfaChiSquared}, where
 $\tau$ must be increased by that amount to maintain fixed $\pfa$
 due to the increasing total noise in $\ze$.
This increase in $\tau$ has to be compensated by an increase in $\energy_{h}$
by that amount to maintain fixed $\pd$.

\incfig
	{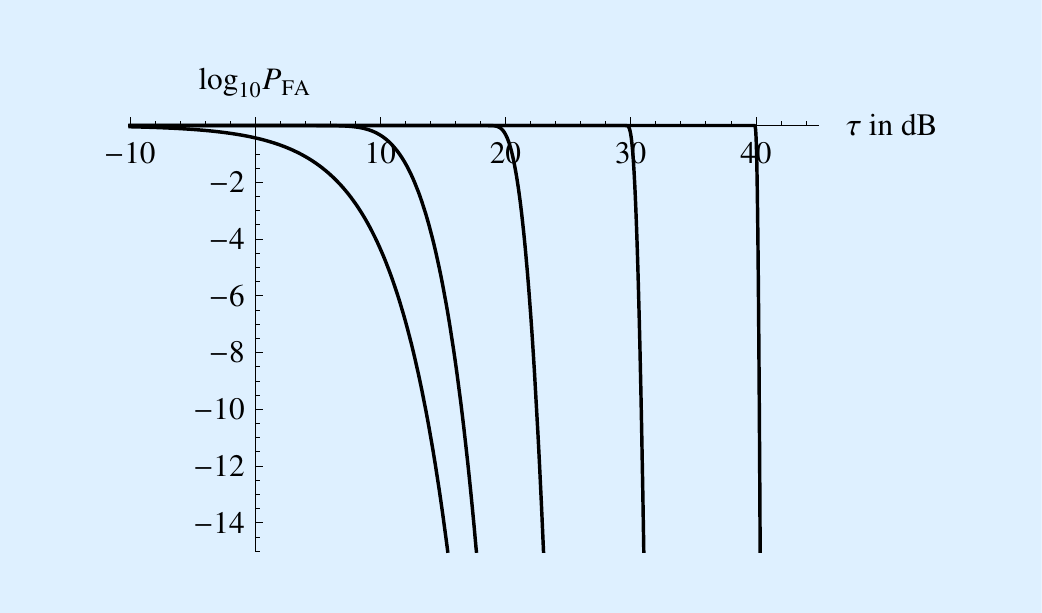}
	{.6}
	{
	The energy detector log-log plot of $\pfa$ against the normalized
	threshold $\tau$ 
	for $K = 10^n$ where $n \in [0,4]$.
	Larger $K$ have higher $\pfa$ for the same $\tau$,
	or require larger $\tau$ for the same $\pfa$.
	Note that the matched filter corresponds to $K=1$, $n=0$,
	and has the most favorable tradeoff of $\tau$ vs $\pfa$.
	}

A major advantage of the ED is that unlike the MF it does not require knowledge of $\esv h$,
but a substantial penalty is paid in sensitivity for $K>1$.
That is one manifestation of a more general principle of signal detection:
The more specific the detector's knowledge of the signal, the better the sensitivity
of a detector that makes use of that knowledge.
Cyclops \citeref{142} prescribed an ED to avoid knowing $\esv h$,
but also prescribed $K \approx 1$ in order to preserve detector sensitivity.
Intuitively the penalty for increasing $K = BT$ in the ED arises because more total
noise energy is captured in $\ze$
either because $B$ is large (noise energy equals noise density times bandwidth)
or $T$ is large (noise energy equals noise power times integration time).
$K \approx 1$ can be accomplished with a carrier-like signal (small
$B \approx 1/T$) or a pulse-like signal (small $T \approx 1/B$).
In both cases the one DOF specifies the phase and amplitude, and thus
this strategy chooses an exact signal waveform $h(t)$
(just as we propose for spread spectrum in Section \ref{sec:pseudonoise}).
Small $B$ was preferred in \citeref{142} to reduce the transmit power for a given $\energy_h$.

\subsection{Radio-frequency interference}

Is there any motivation to increase the signal DOF $K$?
First, the information-bearing nature of the PAM signal of \eqref{eq:pam} is \emph{not} a motivation,
since the only requirement is that \eqref{eq:ott} be satisfied and this only requires that $K = B \, T \ge 1$.
Decreasing $T$ is beneficial because the higher symbol rate implies (all else equal) a higher
information rate, and this in turn requires larger $B$ but not larger $K$.
The primary penalty for increasing the information rate is higher transmit signal power, since
that power is $\energy_{h}/T$, and if $h(t)$ is known and a MF is used in the receiver
we saw in Section \ref{sec:noise} that $\energy_{h}$ does not depend on $B$, $T$, or $K$
for fixed $\pfa$ and $\pd$.
There is, however, one strong motivation for making $K$ large, and that is RFI as we will now see.

Recall from Section \ref{sec:rfi} that the receiver's immunity to RFI is quantified
by the gain $G = FG \times PG$.
If, as is likely, the transmitter designer has no specific knowledge of the RFI
environment at the receiver, he has no opportunity to manipulate $FG$
through the choice of $T$ and $B$ since the temporal and bandwidth characteristics of the RFI are unknown.
However, that designer does have an opportunity to manipulate $PG$ by choosing $K$ to be large.
As long as the receiver has knowledge of $h(t)$ and uses that knowledge to construct
a MF detector, then the transmitter designer can assume that such a large $K$
has no adverse effect on the detection noise sensitivity.
He can therefore focus on choosing $h(t)$ to achieve
\emph{robust} $PG$, meaning the \emph{greatest} $PG$ that
can be \emph{consistently} achieved without
needing to incorporate any specific assumptions about the RFI.

From the transmitter designer perspective, assume the RFI
after the benefit of $FG$ is represented by
a deterministic (but unknown) vector $ \esv r \in \cn{K}$ in 
 the same basis as the signal vector $\esv h$, where $\esv r$ is a unit vector
 that represents the direction of the RFI in $\cn{K}$ and $\energy_{r}$ is
 the energy of the RFI.
If the receiver is presumed to use a MF, then the transmitter designer
knows that the component of RFI at the decision threshold is
$\zm = \energy_{r} \, \hc{\esv h} \esv r$ , and this term
biases the decision in favor of $\energy_{h}>0$ and hence increases $\pfa$.
In choosing $\esv h$ the transmitter thus has an interest in making $\left| \, \hc{\esv h} \esv r \, \right|$
as consistently small as possible.
The factor by which RFI energy is reduced by the MF is
\begin{equation}
\label{eq:interfpgain}
PG = 
\frac{\left\| \, \energy_{r} \, \esv r \, \right\|^{\,2}}{| \, \energy_{r} \, \hc{\esv h} \, \esv r \, |^{\,2}} = \frac{1}{| \, \hc{\esv h} \esv r \, |^{\,2}} \,.
\end{equation}
$PG$ depends on the geometric relationship of two unit vectors, $\esv h$ and $\esv r$:
$PG =1$ when $\esv h$ is colinear with $\esv r$,
$PG = \infty$ when it is orthogonal,
and otherwise $1 < PG < \infty$.

How do we achieve consistently large $PG$?
Two perspectives arrive at similar conclusions.
First, if $\esv h$ is in some sense
spread uniformly throughout $K$ dimensional space,
only a fraction $1/K$
of its unit energy will fall in the direction of any $\esv r$, independent of $\esv r$.
In that case, $PG \approx K$, and there is a clear benefit to making $K$ as large as possible.
Second, consider a basis chosen to include $\esv h$ as the first basis vector.
Then two extreme scenarios are the unit energy of $\esv r$ spread evenly over all
basis vectors, in which case only a fraction $1/K$ of the unit energy is co-linear with 
the one coordinate $\esv h$,
or the unit energy is concentrated in one basis vector, in which case the chance that
this vector happens to be the signal direction is only $1/K$ \citeref{157}.
In either case, $PG \approx K$ in a statistical sense, and
we conclude again that large $K$ is advantageous.
Nevertheless, $PG \approx 1$ can always occur inadvertently when $\esv h \approx \esv r$,
even if that outcome is deemed to be unlikely.

Lacking control over or knowledge of $\esv r$, 
a communication engineer focuses
instead on $\esv h$, which the transmitter designer chooses.
Specifically, the previous argument suggests that he 
should seek a signal which spreads its energy uniformly throughout
the $K$ DOF's.
There is no direct justification for choosing any specific $\esv h$, either for
noise sensitivity or RFI immunity.
Thus, the designer can do no better than
choose $\esv h$ as a random vector $\esv H$ drawn from some appropriate random ensemble.
The transmitter will choose $\esv H$ randomly from this ensemble,
and a receiver knowing $\esv H$ implements a MF using $\esv H$.
The obvious question---how does the receiver know $\esv H$?---
is addressed in Section \ref{sec:pseudorandom}.
With this approach, the designer can answer two useful questions:
First,
$PG = 1/ | \hc{\esv H} \esv r |^{\,2}$ becomes a random variable, and the distribution of $PG$
yields specific information as to how the $PG$ is distributed,
and especially how likely it is that $PG$ will be small
(the RFI mimics the signal)
or large (RFI is highly attenuated by the MF).
Second, the designer can ask how the choice of a random ensemble for $\esv H$ affects the distribution of $PG$,
and in particular if there is any choice that renders that
distribution independent of $\esv r$,
achieving the goal of robust RFI rejection in a statistical sense.

The IP suggests an immediate answer to the second question because
it establishes that the energy of a Gaussian CR $\esv H$ is the same in any direction and thus
 the variance $| \hc{\esv H} \esv r |$ does not depend on $\esv r$.
The only problem is that this ensemble does not incorporate the constraint
that $\| \esv H \| \equiv 1$.
To address this, let $\esv X$ be a Gaussian CR random vector, and choose
$\esv H = \esv X / \| \esv X \|$.
For this choice of $\esv H$, the distribution function of $PG$
in \eqref{eq:interfpgain} is ( \ref{sec:isotropicPrinciple})
\begin{equation}
\label{eq:pgdf}
F_{PG} (p) = \prob{PG \le p} =  \left( 1 - \frac{1}{p} \right)^{K-1} ,\;\; p \ge 1  \,.
\end{equation}
Notably $F_{PG} (p)$ still does not depend on $\esv r$ and
$F_{PG} (p)$ decreases exponentialy with increasing $K$ as illustrated
 in Figure \ref{fig:cdfpgdb}.
\emph{Any} objective for the probability that $PG \le p$ for any $p >0$
can be met by choosing $K$ sufficiently large.
For example, if we require the probability that $PG$ falls below
40 dB to be less than $10^{-10}$,
then we must choose
$K > 230,248$ (e.g. $T = 0.1$ sec and $B = 2.31$ MHz).

\incfig
	{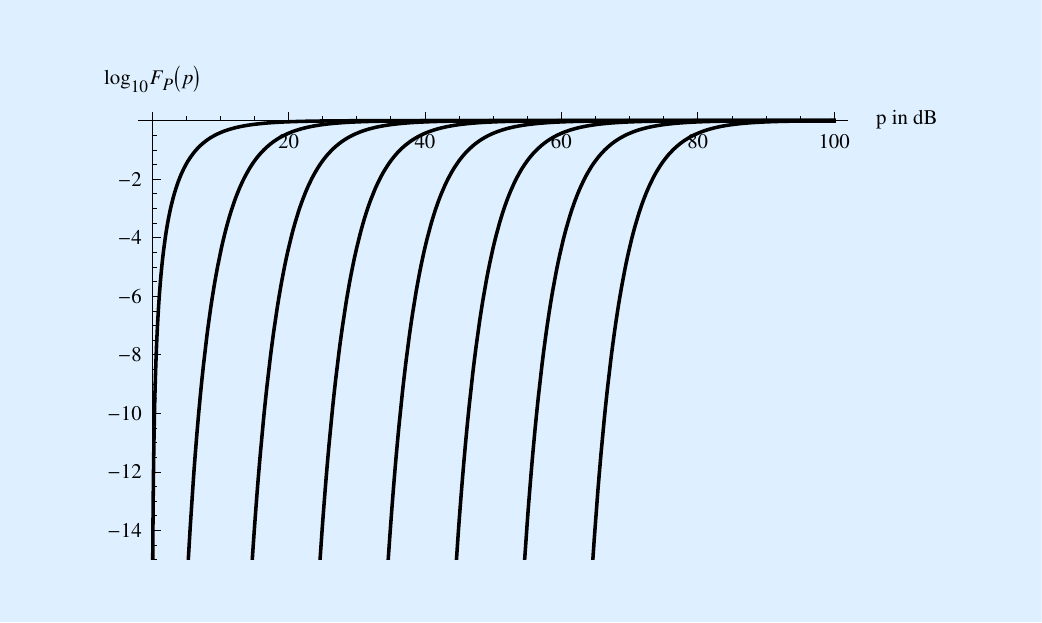}
	{.6}
	{
	The cumulative distribution function $F_{PG} (p)$ of processing gain $PG$ for
	a CR signal $\esv H = \esv X / \| \esv X \|$, a Gaussian CR $\esv X$,
	and any $\esv r$.
	It is plotted on a log-log plot ($10 \log_{10} p$ vs $\log_{10} F_{\pg} (p)$)
	for $K = 10^n$ and $n \in [1,8]$.
	The distribution moves to the right by approximately 10 dB for every order-of-magnitude increase in $K$.
	}

Consider for example a jammer who deliberately tries to
design a worst-case interferer and is aware that $\esv H$
was chosen as described previously, but is not aware of the specific outcome $\esv H$.
This jammer has no basis for choosing a specific $\esv r$;
that is, no choice is better or worse than any other.
There is a duality: Choosing a CR signal $\esv H$ yields a $PG$ distribution
that is independent of $\esv r$, just as a CR noise $\esv N$ results in
a distribution (the same distribution!) that is independent of $\esv h$.
These two statements are compatible, as any randomly chosen $\esv H$ in the former is
entirely suitable for the latter.
For any other way of choosing $\esv H$,
the distribution of $PG$ is dependent on $\esv r$.
That is,
some interferers are rejected more than others, and robust immunity is lost.

There are several conclusions.
First, there are always values of $\esv r \approx \esv H$ for which immunity is poor.
In short, beware of impostors.
However, the probability of this happening inadvertently
can be made arbitrarily small by increasing $K$.
Second,
$K$ can be chosen freely without penalty in noise sensitivity with a MF;
large $K$ is advantageous for RFI immunity and not deleterious to noise immunity.
Third, robustness in RFI immunity requires that $\energy_{h}$
 be spread uniformly over its $K$ DOF's, and this is achieved
if $\esv H$ is Gaussian CR.
Unlike noise immunity, the goal of robust RFI immunity
provides specific information about advantageous
signal characteristics; namely, the signal should
statistically resemble white Gaussian noise.\footnote{
Since a Gaussian CR signal vector remains Gaussian CR after any transformation
of basis, this conclusion is transparent to the basis.
}

Choosing the signal from a CR ensemble in conjunction with the MF
is an equilibrium strategy in the game theoretic sense.
Any deviation from this strategy either degrades the detector sensitivity in noise,
or it abandons robustness in RFI rejection, or both.
Assuming the receiver designer
focuses on the limited information available to the transmitter designer,
each will arrive at this same fundamental conclusion
based on elementary probability theory.
However, turning this principle into a concrete signal design introduces some ambiguity
as discussed next.

\section{CONCRETE SIGNAL DESIGN}
\label{sec:pseudonoise}

Even acknowledging fundamental principles,
obvious questions loom.
How do we choose the signal parameters $B$ and $T$?
Given that the transmitter cannot communicate a chosen signal $\esv h$
to the receiver, can we make a credible argument that $\esv h$
can be guessed by the receiver, or that there are at least a
limited number of options that can be searched?
In addressing these issues, we draw upon extensive experience
in designing terrestrial communication systems.

\subsection{Choosing $T$ and $B$}

The RFI $PG$ depends only on the time-bandwidth product $K= B \, T$,
and thus offers no insight into how a chosen $K$ should be partitioned between
$B$ and $T$.
This design choice must rely on external considerations such as
the transmitter's stellar scanning dwell time and desired data rate
(which influence $T$), impairments in the interstellar medium
(which influences $B$), and the receiver's search strategy.
Generally it can be asserted that $T$ is a scarce resource but $B$ is more ample.
Another consideration is the RFI $FG$ (Section \ref{sec:rfi}),
which depends on the anticipated nature of the RFI.
For persistent narrowband RFI (like other communication signals) it is more advantageous to increase $B$,
but for bursty RFI (like impulsive noise from the electrical grid or motors)
it is more advantageous to increase $T$.
Finally, many current searches
depend entirely on large $FG$ by choosing $K \approx 1$ and thus
perform poorly for RFI that mimics the signal's small time-bandwidth product
but relatively well for other types of interferers.
 
\subsection{Pseudorandom signal}
\label{sec:pseudorandom}

In the random signal approach,
there is no possibility of communicating $\esv h$ to the receiver, 
but the receiver may be able to guess an algorithm used to generate the signal.
This suggests using a \emph{pseudorandom algorithm (PR)} for generating
a signal that is representative of the random ensemble.
The criterion for ''representative" is meeting
standard statistical tests for a sequence that is CR.
The significance or reliability of such tests increases with $K$.
Although the PR algorithm mimics white Gaussian noise,
 the receiver searches for a unique signal 
rather than applying statistical tests to the reception,
and thus there is no extra source of confusion with noise when
the signal is chosen randomly.

\subsection{QPSK signal}
\label{sec:qpsk}

To be successful in acquisition, the receiver must guess both the basis functions
and the PR algorithm.
Of course, in both cases the receiver can search multiple possibilities subject
only to computational power (principally budget and technology) limitations
and must suffer a penalty in $\pfa$.
(There is no possibility of searching over all possibilities, because the $\pfa$
penalty would be too large.)
Based on our terrestrial experience,
either the sampling theorem  \eqref{eq:oseries1} or Fourier \eqref{eq:oseries2} 
basis seems a likely choice.
When resolving the ambiguity of what PR generator to use, Occam's razor --
look for the simplest solution that meets the requirements -- is a good
guiding principle \citeref{583}.
The question then is whether the simplest possible PR generator, one that
mimics a sequence of independent coin tosses, can be used.
The answer is yes if an independent CR (rather than Gaussian CR) $\esv H$ can suffice.
Fortunately, from Table \ref{tbl:flavorIsotropic}
independent CR vectors obey the
energy IP precisely and also approach Gaussian CR in distribution as $K \to \infty$.

The approach used in terrestrial systems is to define a finite constellation\footnote{
This is similar to the constellation of PAM data symbol $\{A _{k} \}$,
but with a completely different motivation.}
from which
components of $\esv H$ are chosen randomly.
The smallest complex-valued constellation for the components of $\esv H$
that can be independent CR and zero mean
has four points
(these properties are inconsistent with a two- or three-point constellation).
This constellation must include the points $(\pm 1 \pm i)$, where 
the real and imaginary components are
independent and chosen by a fair coin toss.
Suppose a binary PR generator statistically mimics a sequence of
$2K$ independent coin tosses $\{ c_{l}, \, 1 \le l \le 2 \, K \}$, where $c_{l} = \pm 1$.
Then the vector
\begin{equation*}
\esv h = [ c_1 + i \cdot c_2 , \; c_3 + i \cdot c_4 , \dots , \; c_{2K-1} + i \cdot c_{2K}]
\end{equation*}
statistically mimics an independent CR random vector $\esv H$, and it is the simplest choice.
This is called a \emph{quaternary phase-shift key} (QPSK) PR generator,
and as expected the distribution of $PG$ for this PR sequence and a Gaussian CR $\esv H$
are virtually indistinguishable as illustrated in Figure \ref{fig:trialsSignal}.

\incfig
	{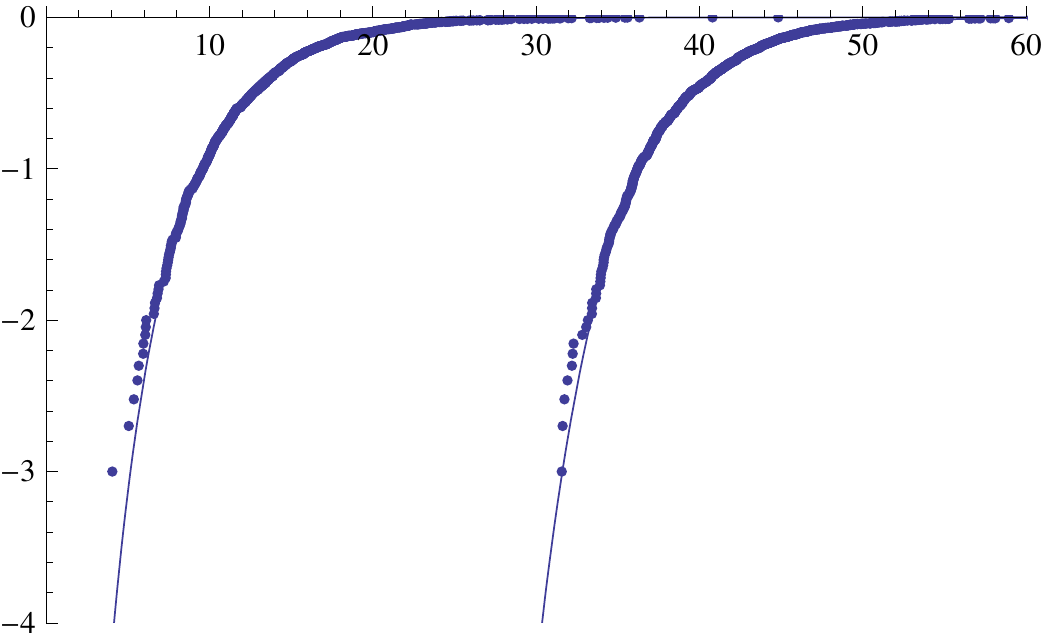}
	{.5}
{The distribution of RFI $P$ in dB for an interferer 
consisting of a 64-QAM digital communication signal with random data symbols, 
one symbol per second, and a 50\% excess bandwidth. 
The two cases shown both assume that $T=10$ seconds, 
with $B = 2$ ($K = 20$) and $B = 1000$ ($K = 10^4$).
Shown are the results of one million trials of a PR QPSK signal
using the RandomTrials[] function of Mathematica.
The vertical axis is $\log_{10}$ of the
fraction of trials for which $P < p$ and the horizontal axis is $p$ in dB.
The solid curve is the theoretical prediction for
a Gaussian CR signal.
}

\subsection{The PR open-ended sequence property}

Another requirement on the PR algorithm is evident upon examining the basis expansions.
The receiver's computational requirements can be dramatically reduced if a mismatch in
DOF $K$ between transmitter and receiver is permitted,
and this is permissible as long as the PR sequence $\{ c_k , \; 1 \le k < \infty \}$
is open-ended and has an unambiguous initial state.
If the bandwidth $B$ remains fixed in \eqref{eq:oseries1} and $c_k$ is mapped
onto coordinates $\esv H$ with DOF $K$ in the transmitter and $M \ne K$ in the
receiver, then detection remains feasible with reduced sensitivity.
If $K > M$ then some transmitted energy is lost, if $K < M$ some unnecessary
noise is added into the signal level estimate, and maximum sensitivity occurs for $M=K$.
The receiver must still search over different values of $B$ (or equivalently different sampling rates), 
but the value of $T$ can be left open.
This required search over $B$ introduces an additional search parameter
and increases the computational burden accordingly.
Similar logic applies to Fourier basis \eqref{eq:oseries2}, except that the search is over $T$
and it is $B$ that is left open.

Any discovery algorithm for an information-bearing signal must search over
starting time and $f_{c}$.
In the discovery of an information-bearing signal
(and spread spectrum is no exception)  there is one additional dimension to search,
which we call a \emph{dilation parameter}.
It can be chosen to be $T$
(equivalent to the frequency sampling or symbol rate $1/T$)
or $B$
(equivalent to the time sampling rate $1/B$).
The most sensitive detectors will search over both these dilation parameters.

\subsection{PR coin-toss generators}

\incfig
	{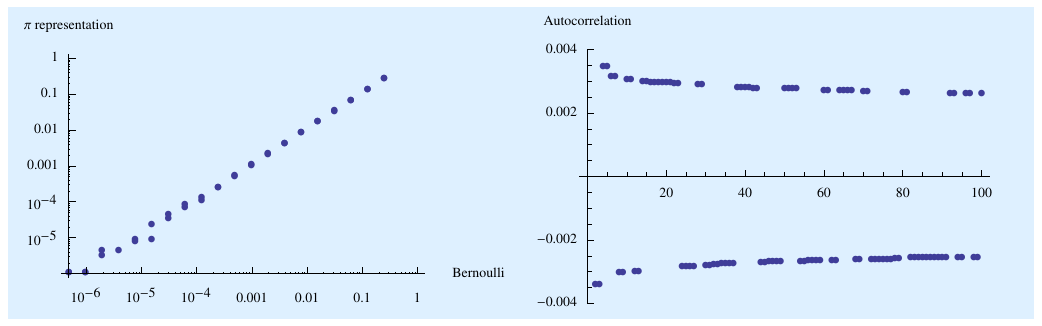}
	{1}
	{
	Tests for independent coin-toss statistics for a two million bit binary expansion of $\pi$.
	On the left,
	the relative frequency of run lengths of $+1$'s and $-1$'s
	against the predicted probabilities.
	The disparities at small relative frequencies is explained by the statistically insignificant number of
	longer sequences.
	On the right,
	the magnitude of the autocorrelation sorted from largest to smallest is plotted with
	the unity autocorrelation at zero lag omitted.
	The predicted value of zero is satisfied
	within a few parts in a thousand.
	}

Because of their explicit coordination,
terrestrial systems do not require the open-ended sequence property
and hence offer little direct guidance as to choice of a PR algorithm.
Nevertheless it is straightforward to find PR algorithms
with this property, including ones that rely on nothing more sophisticated
than the geometry of the square or the circle.
For example,
 common irrational numbers such as
$\pi$, $\sqrt{2}$, and $e$ expanded in base two satisfy the open-ended sequence property and have been observed to satisfy statistical tests for 
independent and identically distributed coin tosses \citeref{600}.
If the receiver has to search over $L$ different algorithms, an increase
in $\energy_{h}$ by $\log L$ is necessary to avoid an increase in
overall $\pfa$ \citeref{572}.

The number $\pi$ seems like a particularly interesting candidate,
especially since it has been known and studied centuries 
and has been confirmed to pass rigorous statistical tests \citeref{601}.
An example of one such test, the frequency of run lengths,
is shown in Figure \ref{fig:statsBernoulli}.
A run of one +1
followed immediately by a -1
should occur close to one quarter of the time, 
two +1's followed by a -1 close to an eighth of the time, etc.
Another test of randomness calculates the sample autocorrelation function.
Since $E[ C_m \cc{C}_n ] = \frac{1}{K} \cdot \delta_{m-n}$ for any white CR vector, at large $K$
the time average autocorrelation should approximate,
\begin{equation*}
\sum c_k \cc{c}_{k-m} \approx \delta_m  \,.
\end{equation*}
Equivalently, the Fourier transform of $c_k$ should be approximately white (constant magnitude).
By the IP,
such signals have their energy spread relatively uniformly over both
a time basis in $t \in [0,T]$, a frequency basis in $f \in [0,B]$, or indeed \emph{any} basis.
Such a signal appears noise-like and stationary, as illustrated by the
$\pi$ representation of Figures \ref{fig:rePi200} and \ref{fig:magPi200}.

\incfig
	{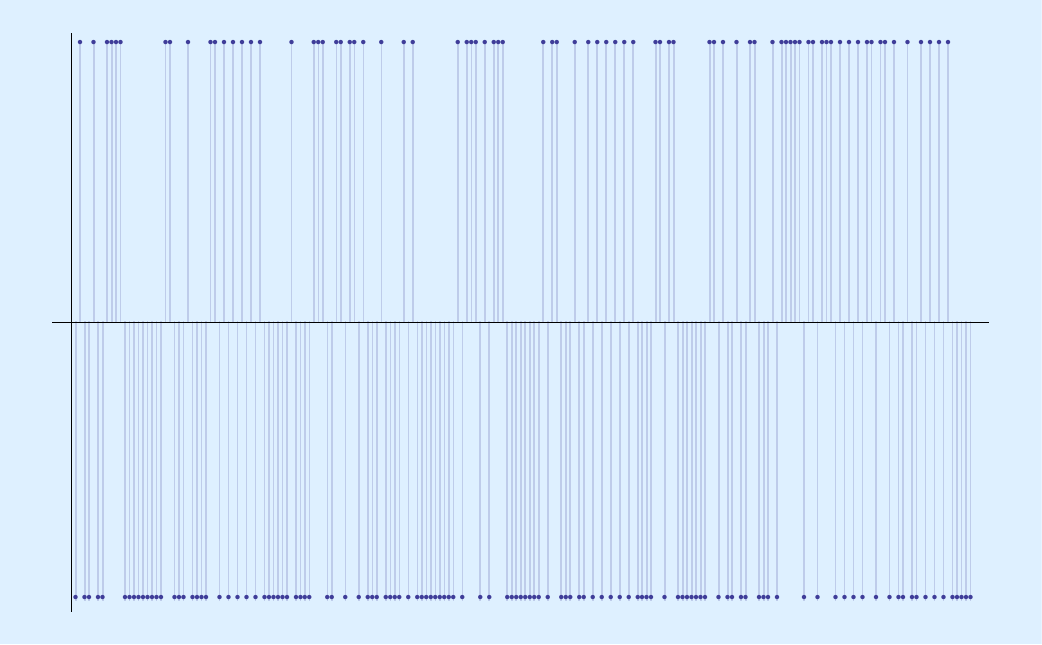}
	{.8}
	{
	Plot of the real part of a PR signal with $K=100$ based on the first 200 bits of the
	binary expansion of $\pi$
	(typically a much larger $K$ would be used).
	By construction,
	the signal energy is spread uniformly over time,
	and is thus relatively immune to pulse-like RFI.
	}

\incfig
	{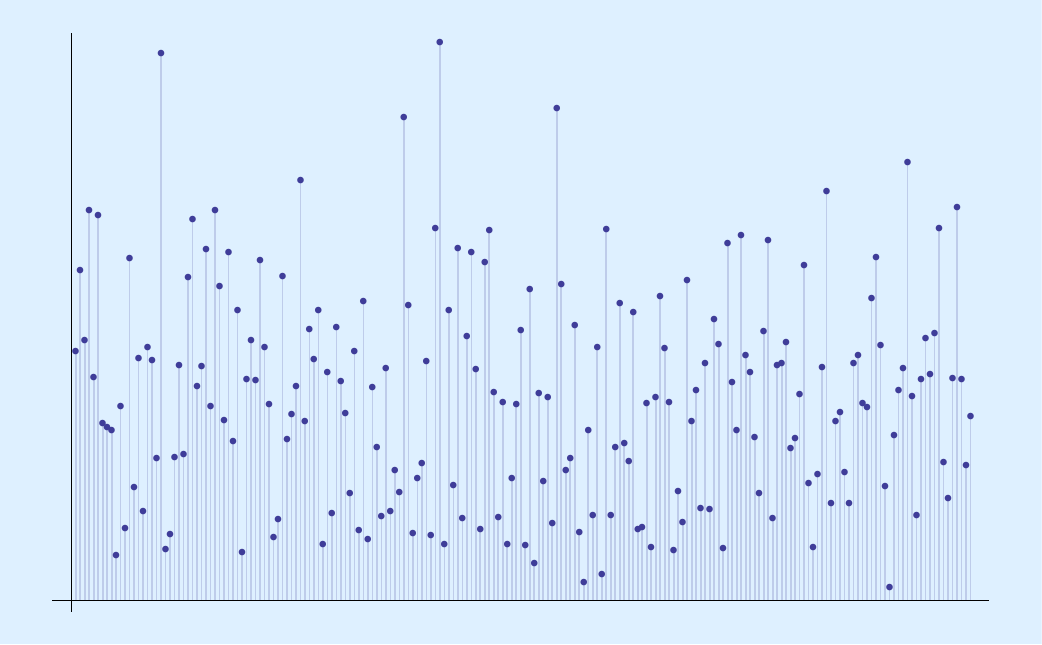}
	{.8}
	{
	Plot of the magnitude of the discrete Fourier transform of the signal in Figure \ref{fig:rePi200}.
	Reflecting the IP, the signal energy is spread relatively uniformly over frequency,
	and is thus relatively immune to narrowband RFI.
	The variation in energy with frequency is consistent with the expected complete randomness.
	}

\section{CONCLUSIONS}

We have addressed the discovery of an information-bearing
signal for interstellar communication in the presence of noise and RFI.
Design for maximum detection sensitivity in noise
offers no insight into the type of signal to transmit or seek.
We have demonstrated that
a signal in which each symbol-period pulse $h(t)$ 
statistically matches a burst of bandlimited white noise
is the only choice that
achieves robust immunity to RFI
(immunity not dependent on the specific
in-band RFI waveform)
for matched filter detection.
This immunity is additive to other forms of RFI mitigation,
and works when RFI is overlapping
in frequency, time, and space.
A signal with this characteristic can be generated by a PR
algorithm that we assert can be guessed by the receiver, such
as a QPSK constellation in conjunction with the binary expansion 
of an irrational number like $\pi$.
This observation informs
 both transmitter and receiver designers as to the desirable
 characteristics of a signal to transmit and to seek.

A consequence of wider signal bandwidth that has not be addressed is
its susceptibility to dispersion and multipath distortion introduced
in the ISM \citeref{582}.
While the same can be said of any information-bearing signal
as the information rate is increased,
spread spectrum increases the severity of this impairment.
This is addressed elsewhere \citeref{572}, where it is shown that
there is a quantifiable no-dispersion region
of $\{ B, \, T, \, f_c \}$ for which these effects are insignificant.
This observation together with a desire to maximize 
$K = B T$ suggests transmitting and searching
for signals near the boundary of the no-dispersion region,
dramatically narrowing the ambiguity in parameters $\{ B, \, T, \, f_c \}$
for both transmitter and receiver.

Other practical issues are yet to be addressed.
What are the consequences of relative motion between transmitter,
receiver, and ISM clouds?
What about the conversion from continuous-time, and the related
granularity in the search space over $\{ f_{c} , B \}$ or $\{ f_{c} , T \}$?
What are the computational requirements for such a search?
How are these parameters related to a stellar scanning strategy?

\section*{ACKNOWLEDGEMENTS}

The author is indebted to
Samantha Blair,
Kent Cullers,
Gerry Harp,
Ian S. Morrison,
Andrew Siemion,
Rick Standahar,
Jill Tarter,
and
Dan Werthimer
for useful discussions and comments.
Peter Fridman provided patient guidance to
the literature on RFI mitigation in radio astronomy.
This research was supported in part by the
National Aeronautics and Space Administration, the SETI Institute,
and the University of California at Berkeley.

\appendix
\addappheadtotoc

\section{INFORMATION-BEARING SIGNALS}
\label{sec:infobear}

Assuming that the information to be conveyed is
digital in nature \citeref{43}, without loss of generality it can
be represented as a sequence of bounded integers
$\{ I_k, \ -\infty < k < \infty \}$ where $k$ represents time
and $I_{k} \in [1,M]$.
The radio electric field is generated by a process of modulation,
turning $\{ I_{k} \}$ into a continuous-time waveform.
In general form, and equivalently at baseband, this waveform is
\begin{equation}
\label{eq:modulation}
y(t) = \sum_{k = -\infty}^{\infty} h_{I_k} (t - k T) \,,
\end{equation}
where the $\{ h_{l} (t) , \, 1 \le l \le M \}$ is a set of $M$
waveforms distinctive in a way that allows the receiver to infer the
corresponding $I_{k}$.
The raw information rate is $R = (\log_2 M) /T$ bits per second.

A passband signal at carrier frequency $f_{c}$ can
readily convey a complex-valued waveform $y(t)$, which can effectively
double or more than double the information rate by allowing a larger $M$.
To see this,
any real-valued passband signal $x(t)$ with carrier frequency $f_{c}$ can be represented in 
terms of an equivalent complex-valued baseband $y(t)$ as
\begin{equation*}
x(t) = \sqrt{2} \cdot  \Re \left\{ y(t) \cdot e^{i 2 \pi f_{c} t} \right\} \,.
\end{equation*}
The $\sqrt{2}$ factor equates the baseband and passband power or energy,
and $\Re$ and $\Im$ denote the real and imaginary parts of a complex number.
From
\begin{equation*}
\sqrt{2} \cdot x(t) \cdot e^{-i 2 \pi f_{c} t}
= y(t) + y^* (t) \cdot e^{-i 4 \pi f_{c} t}
 \,,
\end{equation*}
and $y(t)$ can recovered from $x(t)$ by lowpass filtering  $x(t) \cdot e^{-i 2 \pi f_{c} t}$.

Prominent special cases are orthogonal signaling, in which
the $\{ h_m (t) , \ 1 \le m \le M \}$ are
mutually orthogonal (uncorrelated), and the PAM of \eqref{eq:pam}.
FSK and PSK \citeref{602} are examples of these two signal classes.
Although
the choice of $h(t)$ in PAM is emphasized in this paper,
 the results are readily extended to orthogonal waveforms.
The results of this paper thus apply quite generally to any digital
communication scheme conforming to \eqref{eq:modulation}
(this includes virtually all digital transmission in use terrestrially), 
but not to analog modulation (which has been
prominently applied to some METI experiments \citeref{585}).

\section{ISOTROPIC PRINCIPLE}
\label{sec:isotropicPrinciple}

The statements in Table \ref{tbl:flavorIsotropic} will now be verified.
A zero-mean complex-valued random vector in $\cn{K}$
can be written as $\esv V = \esv A + i \cdot \esv B$.
A full set of second-order statistics
include three $K \times K$ covariance matrices:
$E [ \esv A \trans{\esv A} ]$, $E [ \esv A \trans{\esv B} ]$,
and $E [ \esv B \trans{\esv B} ]$
($\trans{\esv x}$ is the transpose of $\esv x$).
Equivalently we can use two complex-valued matrices
($\hc{\esv x}$ is the conjugate transpose (Hermitian) of $\esv x$)
as \citeref{574}
\begin{equation*}
\esv R_v = E \left[ \esv V \hc{\esv V} \right] \text{   and    }
\tilde{\esv R}_v = E \left[ \esv V \trans{\esv V} \right] \,.
\end{equation*}
$\esv R_v$ is the conventional \emph{covariance matrix}, 
and $\tilde{\esv R}_v$ is a \emph{pseudo-covariance matrix}.
The equivalence can be verified by writing $\esv R_v$ and $\tilde{\esv R}_v$ in terms of $\esv A$ and $\esv B$
and solving for $E [ \esv A \trans{\esv A} ]$, $E [ \esv A \trans{\esv B} ]$,
and $E [ \esv B \trans{\esv B} ]$
in terms of $\esv R_v$ and $\tilde{\esv R}_v$.
In contrast to the real-valued case, the covariance matrix alone does not completely specify
the second-order statistics.
Further, it is readily shown that white CR in Table \ref{tbl:flavorCR}
is satisfied if and only if $\esv R_v = \sigma^{\,2} \cdot \esv I$
and $\tilde{\esv R}_v = \esv 0$.

Let $\esv U$ be a unitary $K \times K$ matrix
($\hc{\esv U} \, \esv U  = \esv U \, \hc{\esv U} = \esv I$)
corresponding to a transformation from one orthonormal basis to another.
For example $\esv U$ is the discrete Fourier transform (DFT) when converting from time
to Fourier basis.
If $\esv V$ is white CR, then so too is $\esv U  \esv V$, as verified by
\begin{align*}
&E \left[ ( \esv U \esv V ) \hc{( \esv U \esv V )} \right] 
=  \esv U \, \left( E \left[ \esv V  \hc{\esv V} \right] \, \right) \hc{\esv U} 
= \sigma^{\,2} \cdot \esv U \hc{\esv U} = \sigma^{\,2} \cdot \esv I \\
&E \left[ ( \esv U \esv V ) \trans{( \esv U \esv V )} \right] 
= E \left[  \esv U \esv V  \trans{\esv V} \trans{\esv U} \right]
=  \esv U \, \left( E \left[ \esv V  \trans{\esv V} \right] \, \right) \trans{\esv U} 
= \esv 0 \,.
\end{align*}
Since $\esv U  \esv V$ is Gaussian when $\esv V$ is Gaussian,
 a Gaussian CR vector remains Gaussian CR after coordinate transformation.
 
Consider the distribution of $\zm$ and $\ze$ given by \eqref{eq:dt}.
By the IP, $\hc{\esv h}\,\esv N_1$ is Gaussian with variance $\sigma^{\,2}$
not dependent on $\esv h$.
For any $K \times K$ unitary $\esv U$, $e^{-i \theta} \cdot \esv U$ is also unitary
and thus (since a unitary transformation is norm-preserving)
\begin{equation*}
\ze =
\left\| \sqrt{\energy_h}  \cdot \esv U \esv h +   \esv N_2 \right\|
\end{equation*}
where by the IP since
$\esv N_1$ is Gaussian CR then so too is $\esv N_2 = e^{-i \theta} \esv U \esv N_1$.
In particular choose any $\esv U$ such that $\esv U \esv h = \trans{[1 \, 0 \, 0 \dots 0 ]}$
($\esv h$ is the first column of $\hc{\esv U}$, or equivalently the first basis vector).
The distribution of $\ze$ does not depend on $\esv h$ because the
distribution of $\esv N_2$ does not depend on $\esv U$.
For $\esv h = \esv 0$
\begin{equation*}
\pfa = \prob{Z_{E} > \lambda} = 
\prob{\sum_{i=1}^{K} \, \frac{| n_{i} |^{2} }{\sigma^{2}/2} > \frac{2 \lambda^{2}}{\sigma^{2}}} \,,
\end{equation*}
where the left side is $\chi^{2}(2 K)$, and $\pfa$ can be calculated from the $\chi^{2}$ distribution function
as shown in Figure \ref{fig:pfaChiSquared}.
Similarly, $\zm$ is $\chi^{2}(2)$ and thus has the same distribution as $\ze$ for $K = 1$.

Our interest is in the random variable $PG$ in \eqref{eq:interfpgain}
when $\esv h$ is a random vector $\esv H = \esv X / \| \esv X \|$,
and $\esv X$ is Gaussian CR.
Let $\esv U$ be any unitary matrix whose first column is $\esv r$.
By the  IP, $\esv V = \hc{\esv U} \, \esv X = \trans{[ V_1, \, V_2, \dots , \, V_K ]}$ is also
Gaussian CR.
Then from the norm preserving property $PG$ is a function of $\esv r$
but its distribution is not,
\begin{equation*}
PG = \frac{\left\| \, \esv X \, \right\|^{\,2}}
	{\left| \, \hc{\esv X} \, \esv r \, \right|^{\,2}} 
= \frac{\left\| \, \hc{\esv U} \, \esv X \, \right\|^{\,2}}
	{\left| \, \hc{\esv X} \, \esv U \hc{\esv U} \, \esv r \, \right|^{\,2}}
= \frac{\left\| \, \esv V \, \right\|^{\,2}}
	{\left| \, \hc{\esv V} \, \hc{\esv U} \, \esv r \, \right|^{\,2}}
= \frac{\left\| \, \esv V \, \right\|^{\,2}}
	{\left| V_1 \, \right|^{\,2}} \,.
\end{equation*}
Letting $W = \sum_{k=2}^{K} \left| V_k \right|^{\,2}$ , the distribution function is
\begin{equation*}
\prob{PG \le p} =
\prob{W \le (p-1) \cdot \left| V_1 \right|^{\,2}}
= \int_{0}^{\infty} \prob{w \le (p-1) \cdot \left| V_1 \right|^{\,2}} f_W (w) \, dw
\end{equation*}
where $W$ is $\chi^{\,2} (2K-2)$ and $ \left| V_1 \right|^{\,2}$ is $\chi^{\,2} (2)$.
Evaluating the integral yields \eqref{eq:pgdf}.

The asymptotic property for an independent CR vector
mentioned in Table \ref{tbl:flavorIsotropic} is significant for characterizing $PG$
when $\esv H$ is independent CR but not Gaussian (e.g. a QPSK signal
drawn from independent coin-tosses).
The distribution of $\hc{\esv X} \esv r$
will approach Gaussian as $K \to \infty$ for almost all values of $\esv r$
since
\begin{equation*}
\hc{\esv X} \, \esv r = \sum_{k=0}^{K-1} \cc{X}_k \, r_k 
\end{equation*}
is a sum of independent random variables.
Although the terms in the summation are not identically distributed,
the Lyapunov version of the Central Limit Theorem \citeref{195}
asserts that $\Re \{ \hc{\esv X} r \}$ 
and $\Im \{ \hc{\esv X} r \}$
approach Gaussian in distribution as $K \to \infty$
if at most a finite number of the $\{ r_k , \; 0 \le k < \infty \}$ 
are zero and the third moments of $\Re \{ X_k, \; 0 \le k < \infty \} $ and 
$\Im \{ X_k , \; 0 \le k < \infty  \}$ obey certain upper bounds.
For the QPSK PR generator of Section \ref{sec:qpsk} these third moments are identically zero,
so relatively rapid convergence to Gaussian is assured.
Could any distributions other than Gaussian
result from the sum of independent random variables?
Yes, the larger class of stable distributions \citeref{190};
however, the only stable distribution with finite variance
(and $E | \hc{\esv X} r |^{\,2} = \sigma^{\,2} < \infty$) is the Gaussian.

\section{NOISE STATISTICS}
\label{sec:noiseStats}

A noise vector $\esv N$ that originates as white Gaussian noise
on the passband channel is Gaussian CR \citeref{43}
as long as $f_c > B$.
$\esv N$ is Gaussian because the conversion is linear.
To demonstrate that $\esv N$ is also white CR, choose $K$ 
orthonormal baseband basis functions
$\{ g_k (t) , \, 1 \le k \le K \}$ on $t \in [0,T]$, and assume
real-valued white Gaussian noise $M(t)$ at passband with autocorrelation 
$E[ M(t) M(s) ] = N_0 \cdot \delta(t-s)$.
Then
\begin{align}
\notag
&N_k = \sqrt{2} \cdot  \int_0^T M(t) e^{- i 2 \pi f_{c} t} \cc{g}_k (t) \, \text d t \\
\label{eq:rkm}
&E \left[ N_k N_m \right] = 2 N_0 \int_0^T e^{- i 4 \pi f_{c} t} \cc{g}_k (t) \cc{g}_m (t) \, \text d t = 0 \ \ \ \text{for} \ \ \  f_{c} > B  \\
\notag
&E \left[ N_k \cc{N}_m \right] = 2 N_0 \int_0^T  \cc{g}_k (t) g_m (t) \, \text d u = 
2 N_0 \cdot \delta_{k-m} \,.
\end{align}
Note that $\cc{g}_k (t) \cc{g}_m (t)$
falls in frequency band $f \in [0,2B]$, and \eqref{eq:rkm} is a
Fourier transform evaluated at frequency $2 f_{c}$, which evaluates to
zero as long as $2 f_c > 2 B$.

\section*{REFERENCES}

\addcontentsline{toc}{section}{REFERENCES}


\bibliographystyle{ieeetr}

\footnotesize
\small
\bibliography{SETI-references}
 

\normalsize
\include{Authors}

\end{document}

%% file: SETI-headers.tex
\usepackage{geometry}                
\usepackage[parfill]{parskip}    
\usepackage{graphicx}
\usepackage{amssymb}
\usepackage{amsmath}
\usepackage{amsthm}
\usepackage{url} 
\usepackage{lineno} 
\usepackage{endnotes} 
\usepackage{longtable} 
\usepackage{booktabs}
\usepackage[font=small,labelfont=bf]{caption} 
\usepackage{graphicx}
\usepackage{color} 
\usepackage{datetime} 

\usepackage{appendix}

\newcommand	
	{\incfig}	
	[3]	
	{
	\begin{figure}
		\begin{center}
			\includegraphics
				[width=#2\textwidth]	
				{fig-#1}	
			\caption
				{#3}
			\label
				{fig:#1}	
		\end{center}
	\end{figure}
	}

\newcommand 
	{\doctable}
	[5] 
	{
	\begin{table}[t]
	\centering
	#1
	\caption{#3}
	\label{tbl:#2}
	\medskip
	\begin{tabular}{#4}
	#5
	\end{tabular}
	\end{table}
	}

\newcommand 
	{\citeref}
	[1] 
	{\cite{RefWorks:#1}}
	
\newcommand 
	{\citereftwo}
	[2] 
	{\cite{RefWorks:#1, RefWorks:#2}}

\newcommand 
	{\citerefthree}
	[3] 
	{\cite{RefWorks:#1, RefWorks:#2, RefWorks:#3}}
	
\newcommand 
	{\citereffour}
	[4] 
	{\cite{RefWorks:#1, RefWorks:#2, RefWorks:#3, RefWorks:#4}}

\newcommand 
	{\citereffive}
	[5] 
	{\cite{RefWorks:#1, RefWorks:#2, RefWorks:#3, RefWorks:#4,RefWorks:#5}}
	
\newcommand 
	{\citerefsix}
	[6] 
	{\cite{RefWorks:#1, RefWorks:#2, RefWorks:#3, RefWorks:#4,RefWorks:#5,RefWorks:#6}}

\newcommand{\trans}[1]{{#1}^{\ensuremath{\mathsf{T}}}}           
\newcommand{\cc}[1]{{#1}^{\ast}}                            			
\newcommand{\hc}[1]{{#1}^{\dagger}}                            		
\newcommand{\real}[1]{\Re \{ #1 \}}						
\newcommand{\imag}[1]{\Im \{ #1 \}}						

\newcommand{\prob}[1]{ \text{Pr} \left\{ #1 \right\} }			

\newcommand{\esv}[1]{\mathbf{#1}}						
\newcommand{\field}[1]{\mathbb{#1}}					
\newcommand{\cn}[1]{\field{C}^#1}						
\newcommand{\energy}{\mathcal E}						

\newcommand{\power}{\mathcal P}						
\newcommand{\pg}{\mathcal P}							

\definecolor{color1} {rgb}{.8,0,0}
\definecolor{color2} {rgb}{0,0,.8}
\definecolor{color3} {rgb}{0,.8,0}
\definecolor{blackANDwhite}{rgb}{0,0,0}



%% file: Authors.tex
\section*{AUTHOR}
\addcontentsline{toc}{section}{AUTHOR}
 
David G. Messerschmitt is the Roger A. Strauch Professor Emeritus of Electrical Engineering and Computer Sciences (EECS) at the University of California at Berkeley. At Berkeley he has previously served as the Interim Dean of the School of Information and Chair of EECS. He is the co-author of five books, including \emph{Digital Communication} (Kluwer Academic Publishers, Third Edition, 2004). He served on the NSF Blue Ribbon Panel on Cyberinfrastructure and co-chaired a National Research Council (NRC) study on the future of information technology research. His doctorate in Computer, Information, and Control Engineering is from the University of Michigan, and he is a Fellow of the IEEE, a Member of the National Academy of Engineering, and a recipient of the IEEE Alexander Graham Bell Medal recognizing ``exceptional contributions to the advancement of communication sciences and engineering''.